\documentclass[physics,research paper,accept,oneauthor,pdftex]{Definitions/mdpi}
\usepackage{graphicx,amsmath,amssymb,amsfonts,latexsym,color,dcolumn,bm}
\usepackage{revsymb}
\usepackage{gensymb}
\usepackage{subcaption}
\setitemize{parsep=6pt,itemsep=0pt,leftmargin=*,labelsep=5.5mm}
\setenumerate{parsep=6pt,itemsep=0pt,leftmargin=*,labelsep=5.5mm}
\setlist[description]{itemsep=0mm}   
\usepackage[compat=1.1.0]{tikz-feynman}
\usepackage{subcaption}
\newcommand{\be}{\begin{equation}}
\newcommand{\ee}{\end{equation}}
\newcommand{\bea}{\begin{eqnarray}}
\newcommand{\eea}{\end{eqnarray}}
\newcommand{\ra}{\rangle}
\newcommand{\la}{\langle}

\usepackage[utf8]{inputenc}
\usepackage{gensymb}
\usepackage{braket}
\usepackage[utf8]{inputenc}
\usepackage{caption}
\usepackage{subcaption}
\usepackage{geometry}
\usepackage{array}
\usepackage{comment}

\usepackage{revsymb}
\usepackage{gensymb}
\usepackage{array}
\usepackage{gensymb}
\usepackage{braket}
\usepackage[utf8]{inputenc}
\usepackage{caption}
\usepackage{subcaption}
\usepackage{geometry}
\usepackage{comment}
\usepackage{url}
\usepackage{wrapfig}
\usepackage{url}
\usepackage{enumitem}

\usepackage[colorinlistoftodos]{todonotes}

\usepackage{caption}
\usepackage{subcaption}
\firstpage{1} 
\pubvolume{00}
\issuenum{0}
\articlenumber{0}
\pubyear{2022}
\copyrightyear{2022}
\history{Received: 22 April 2022; Published: 19 October 2022; Updated: 30 May, 2023}
\updates{yes} 

\Title{New Insights into the Lamb Shift: The Spectral Density of the Shift}

\Author{{G. Jordan Maclay}}

\AuthorNames{G. Jordan Maclay}

\address[1]{%
{Quantum Fields LLC}, 147 Hunt Club Drive,  St.  Charles, IL 60174, USA; Emeritus, University of Illinois at Chicago, Chicago, IL 60607  jordanmaclay@quantumfields.com}

\corres{Correspondence: jordanmaclay@quantumfields.com}

\abstract{In an atom, the interaction of a bound electron with the vacuum fluctuations of the electromagnetic field leads to complex shifts in the energy levels of the electron, with the real part of the shift corresponding to a shift in the energy level and the imaginary part to the width of the energy level. The most celebrated radiative shift is the Lamb shift between the $2S_{1/2}$ and the $2P_{1/2}$ levels of the hydrogen atom.~The measurement of this shift in 1947 by Willis Lamb Jr. proved that the prediction by Dirac theory that the energy levels were degenerate was incorrect. Hans~Bethe's calculation of the shift demonstrated the renormalization process required to deal with the divergences plaguing the existing theories and led to the understanding that it was essential for theory to include interactions with the zero-point quantum vacuum field. This was the birth of modern quantum electrodynamics (QED).  Other calculations of the Lamb shift followed by Welton and Power in an effort to clarify the physical mechanisms leading to the shift. We have done a calculation of the shift using a group theoretical approach which gives the shift as an integral over frequency of a function, which we call the spectral density of the shift. The spectral density reveals how different frequencies contribute to the total energy shift.  We find, for example, that half the radiative shift for the ground state 1S level in H comes from photon energies below 9700 eV, and that the expressions by Power and Welton do not have the correct low frequency behavior, although they do give approximately the correct value for the total shift.} 


\begin{document}


\orcidA{0000-0003-4901-1942}
\section{Introduction}

In astronomy, in quantum theory, in quantum electrodynamics (QED), there have been periods of great progress in which solutions to challenging problems have been obtained, and the fields have moved forward.  However, in some cases getting the right answers can still leave fundamental questions unanswered. The Big Bang explained the origin of the cosmic background radiation, but left the problem of why the universe appears to be made of matter and not equal amounts of matter and antimatter\cite{astro}. In quantum theory, we can compute the behavior of atoms yet we cannot describe a measurement in a self-consistent way, or make sense of the collapse of a photon wavefunction from a near infinite volume to a point\cite{espeg}. In quantum electrodynamics we can compute the Lamb shift of the H atom to 15 decimal places\cite{beyer}, yet we are left with the paradox of using perturbation theory to remove infinite terms, or to understand a quantum vacuum with infinite energy.  In this paper, we examine different approaches to the computation of the non-relativistic Lamb shift. For all these approaches, the Lamb shift can be expressed in different ways as an integral over frequency of a spectral density. We analyze the differences in the spectral densities for the different approaches as a function of frequency and compare the spectral densities to those obtained by using a group theoretical analysis. The integral of the spectral density over all frequencies gives the corresponding value of the Lamb shift.

Feynman called the the three page long 1947 non-relativistic Lamb shift calculation by Hans Bethe the most important calculation in quantum electrodynamics because it tamed the infinities plaguing earlier attempts. When the sum over all states is evaluated numerically, it gives a finite prediction that agreed with experiment\cite{lam1}\cite{bethe}.  Dirac said it "fundamentally changed the nature of theoretical physics." Yet when this calculation is explored more deeply, questions arise about it and about other calculations of the Lamb shift, for example those by Welton \cite{welton} and Power\cite{power}, that employ different methods that have different low frequency behavior from Bethe's result yet give approximately the same value for the level shift \cite{mil}.  These three approaches to the Lamb shift and the corresponding vacuum energy densities have also been considered in \cite{passante}.  

There is an intimate relationship between radiative shifts and vacuum fluctuations. The shift can be interpreted as arising from virtual transitions induced by the quantum fluctuations of the electromagnetic field. Since the vacuum field contains all frequencies, virtual transitions to all states, bound and scattering, are possible.  These short lived virtual transition result in a slight shift in the average energy of the atom, a shift which we call the Lamb shift \cite{gjmradiative}. We note that the Lamb shift can also be described as an interaction of the electron with its own radiation field, yielding the same results as the vacuum field\cite{mil}.

Bethe's calculation was based on second order perturbation theory applied to the minimal coupling of the atom with the vacuum field $(e/mc)\boldsymbol{A \cdot p}$ and a dipole approximation. This interaction leads to the emission and absorption of virtual photons corresponding to virtual transitions. The shift is expressed as a sum over the intermediate states reached by virtual transitions. The predicted shift is divergent, but Bethe subtracted the term that corresponded to the linearly divergent vacuum energy shift for a free bare electron, essentially doing a mass renormalization to remove this higher order divergence in the spectral density for the shift. For S states, the resulting spectral density has a 1/frequency behavior for high frequencies giving a logarithmic divergence in the shift.

Welton's model for computing the Lamb shift was based on the perturbation of the motion of a bound electron in the H atom due to the quantum vacuum fluctuations altering the location of the electron, which resulted in a slight shift of the bound state energy \cite{welton}\cite{mil}\cite{gjmradiative}. This simplified intuitive model predicts a spectral density proportional to 1/frequency for all frequencies and a shift only for S states. The approach of Feynman\cite{feyn}, interpreted by Power \cite{power}, considers a large box containing H atoms and is based on the shift in the energy in the quantum vacuum field due to the change in the index of refraction arising from the presence of H atoms.  This approach predicts that the shift in the energy in the vacuum field around the H atoms exactly equals the radiative shift predicted by Bethe for all energy \cite{mil}\cite{passante}. It gives a spectral density with the same high frequency dependence as Bethe, but a different low frequency dependence. A similar calculation to Power's models the Lamb shift as a Stark shift \cite{mil}.

The Lamb shift has been previously computed using O(4) symmetry \cite{lieber} and by a different approach from ours using SO(4,2) symmetry \cite{huff}. We present the results of a calculation of the Lamb shift that is based on a SO(4,2) group theoretical analysis of the H atom that allows us to determine the dependence of the shift on frequency with no sum over states\cite{gjmdynam}. The degeneracy group of the non-relativistic H atom is O(4), with generators angular momentum operator $\boldsymbol{L}$ and Runge-Lenz vector $\boldsymbol{A}$. A representation of O(4) of dimension $n^2$ exists for each value of the principal quantum number $n$, where the angular momentum $L$ has values from 0 to $n-1$, and there are $2L +1$ possible values of $L_z=m$. If we extend this group by adding a 4 vector of generators we get the non-invariance group SO(4,1) which has representations that include all states of different $n$ and $L$ and operators that connect states with different principal quantum numbers. Adding a 5 vector of additional generators gives the group SO(4,2) and allows us to express Schrodinger's equation in terms of the new generators, and to make effective group theoretical calculations \cite{gjmdynam}. We use basis states that allow us to include both bound and scattering states seamlessly \cite{brow} and no sum over states appears in the final expression for the spectral density. One advantage of this approach is that for each energy level we can easily compute a spectral density for the shift whose integral over frequency from 0 to $mc^2/\hbar$ is the radiative shift that includes transitions to all possible states. Thus we can see how different frequencies of the vacuum field contribute to the radiative shift. 

We compare the different approaches of Bethe, Welton and Power to the group theoretical spectral density of the non-relativistic Lamb shift for the 1S ground state, the 2S and 2P levels. With this new picture of the Lamb shift, we have found differences between the various approaches. Knowing the spectral density of the shift provides new insights into understanding the Lamb shift.

\section{Background of Radiative Shift Calculations}

The first calculation of the Lamb shift of a hydrogen atom was done by Bethe in 1947, who assumed the shift was do the interaction of the atom with the vacuum field.  He calculated the shift using second order perturbation theory, assuming that there was minimal coupling in the Hamiltonian:
\be
H_{int}=-\frac{e}{mc}\textbf{A} \cdot \textbf{p}+
\frac{e^2}{2mc^2} \textbf{A}^2
\ee
where $m$, $e$, and $\textbf{p}$ are the mass, charge and momentum of the electron, $c$ is the speed of light in vacuum, and $\textbf{A}$ is the vector potential in the dipole approximation for the vacuum field in a large quantization volume $V$ 
\be
\textbf{A}=\sum_{\textbf{k},\lambda}(\frac{2\pi\hbar c^2}{\omega_k V})^{1/2}(a_{\textbf{k},\lambda}+a^{\dagger} _{\textbf{k},\lambda})\textbf{e}_{\textbf{k}\lambda}
\ee
where the sum is over the virtual photon wave number $\mathbf{k}$, where $kc=\hbar \omega_{k}$, the energy of the virtual photon, and the polarization $\lambda$; $a_{\textbf{k},\lambda}$ and $a^{\dagger} _{\textbf{k},\lambda}$ are the annihilation and creation operators, and $\textbf{e}_{\textbf{k}\lambda}$ is a unit vector in the direction of polarization of the electric field.  The shift from the $\textbf{A}^2$ term is independent of the state of the atom and is therefore neglected.  The total shift $\Delta E_{nTot}$ for energy level n of the atom in state $|n\ra$ is given by second order perturbation theory as \cite{mil} 
\be
\Delta E_{nTot}=-\frac{2}{3\pi}\frac{\alpha}{m^2c^2}\sum_m |\textbf{p}_{mn}|^2 \int \frac{EdE}{E_m - E_n+E}
\ee
where the integral is over the quantum vacuum field energy $E=\hbar \omega$ and the momentum matrix elements are $|\textbf{p}_{mn}|=|\la m|\textbf{p}|n\ra|$. The sum is over all intermediate states $|m \ra$, scattering and bound, where $m\neq n$. The fine structure constant is $\alpha=e^2/\hbar c$.

The integrand in Eq. 3 has a linear divergence. Bethe observed that this divergence in Eq. 3 corresponded to the integral that occurs when the binding energy vanishes $(E_m-E_n)\rightarrow 0$ and the electrons are free:


\be
\Delta E_{free}= -\frac{2}{3\pi}\frac{\alpha}{m^2c^2}\sum_m |\textbf{p}_{mn}|^2 \int dE.
\ee
He subtracted this divergent term $\Delta E_{free}$ from the total shift $\Delta E_{nTot}$ 
\be
\Delta E_{nL}=\Delta E_{nTot}-\Delta E_{free}
\ee
to obtain a finite observable shift $\Delta E_{nL}$ for the state $|nL\ra$  
\be
\Delta E_{nL} = \frac{2\alpha}{3\pi (mc)^2}\sum_m ^s |\textbf{p}_{mn}|^2 \int_0^{ \hbar\omega_c} dE \frac{(E_m-E_n)}{E_m - E_n + E -i\epsilon}  ,
\ee
where $\omega_C$ is a cutoff frequency for the integration that Bethe took as $\hbar \omega_c = mc^2$.
Using an idea from Kramers, Bethe did this renormalization, taking the difference between the terms with a potential present and without a potential present, essentially performing the free electron mass renormalization. He reasoned that relativistic retardation could be neglected and the radiative shift could be reasonable approximated using a non-relativistic approach and he cut the integration off at an energy corresponding to the mass of the electron.  He obtained a finite result that required a numerical calculation over all states, bound and scattering, that gave good agreement with measurements \cite{lam1}\cite{bethe}\cite{bandsbook}. 

The spectral density in the Bethe formalism, which we will analyse, is the quantity in Eq. 6 being integrated over $E$. It includes the sum over states $m$. The term for $m$ represents the contribution to the Lamb shift for the virtual transition from state $n$ to state $m$. Note since the ground state is the lowest state, all intermediate states have higher energies so the ground state shift has to be positive. 

For the purposes of comparison to the other calculations of the Lamb shift it is helpful to show the next steps Bethe took to evaluate the shift $\Delta E_{n}$ for S states, which have the largest shifts. Note that the spectral density we will analyse in Eq. 6 is not affected by the subsequent approximations Bethe made to evaluate the integral. First the E integration is done:
\be
\Delta E_n^{Bethe}=\frac{2\alpha}{3\pi}(\frac{1}{mc})^{2}\sum_{m}|\textbf{p}_{nm}|^{2}(E_m-E_n)ln \frac{(mc^2+E_m-E_n)}{|E_m-E_n|} .  
\ee
To simplify the evaluation Bethe assumed $|E_m-E_n|<< mc^2$ in the logarithm and that the logarithm would vary slowly with $m$ so it could be replaced by an average value
\be
\widehat{\Delta {E}}_{n}^{Bethe}=\frac{2\alpha}{3\pi}(\frac{1}{mc})^{2}ln \frac{mc^2}{|E_m-E_n|_{Ave}} \sum_{m}|\textbf{p}_{nm}|^{2}(E_m-E_n)
\ee
where the hat over the $\Delta E$ indicates this is an approximation to Eq. 7.  Now that the E integration is done, the spectral density is no longer manifest. The summation can be evaluated using the dipole sum rule
\be
2 \sum_{m}^{s}|\textbf{p}_{nm}|^2\left(E_{m}-E_{n}\right)=\hbar^2 \left\langle n\left|\nabla^{2} V\right| n\right\rangle.
\ee
The value of the Laplacian with a Coulomb potential V=$-Ze^2/r$ is $\nabla^{2} V(r)=4 \pi Z e^2\delta(\textbf{r})$ so we have 
\be
\left\langle n\left|\nabla^{2} V\right| n\right\rangle=4 \pi Z e^2|\psi_n(0)|^2,  
\ee
where $\psi(r)$ is the wave function for a Coulomb potential
and $|\psi_n(0)|^2$ is zero except for $S$ states 
\be
|\psi_{n}(0)|^{2}=\frac{1}{\pi}\big( \frac{Z\alpha mc}{n\hbar}\big)^{3}.
\ee 
For S states, this gives an energy shift equal to \cite{mil}:
\begin{equation}
  \widehat{\Delta {E}}_{n}^{Bethe} =  \frac{4mc^{2}}{3\pi}\alpha (Z\alpha)^{4}\frac{1}{n^3}ln \frac{mc^2}{|E_m-E_n|_{Ave}}.
\end{equation}
where the so called Bethe log for an S states with principal quantum number n is 
\be
ln \frac{mc^2}{|E_m-E_n|_{Ave}}=\frac{\sum_{m}|\textbf{p}_{nm}|^{2}(E_m-E_n)ln \frac{mc^2}{|E_m-E_n|}}{ \sum_{m}|\textbf{p}_{nm}|^2\left(E_{m}-E_{n}\right)}  
\ee
where the sum is over all states, bound and scattering. Bethe also has extended the formalism for shifts for states that are not S states \cite{bandsbook}.

Regarding the approximations Bethe made to obtain Eq. 8 from Eq. 7 and the use of the Bethe log Eq. 13, he commented: "The important values of $|E_m-E_n|$ will be of order of the ground state binding energy for a hydrogenic atom. This energy is very small compared to $mc^2$ so the log [in our Eq. 7] is very large and not sensitive to the exact value of $(E_m-E_n)$. In the numerator we neglect $(E_m-E_n)$ altogether and replace it by an average energy \cite{bandsbook}." Our work shows that Bethe was correct that the relative contribution from energies of the order of the ground state is very important, but we find the contribution from higher energy scattering states is very significant, and therefore that the approximation $|E_m-E_n|<<mc^2$ is not valid for higher energy scattering states for which $E_m$ increases to the value $mc^2$. We are not aware of any quantitative estimates of the error in the approximation.  The difference, $0.3\%$, between our value for the total 1S shift and that of Bethe may be due to this approximation, although we have not verified this.  On the other hand Bethe's approximation may have made his non-relativistic approach viable.

To provide a more intuitive physical picture of the shift, Welton considered the effect of a zero-point vacuum field on the motion of an electron bound in a coulomb potential $V(\textbf{r})$ at a location $\textbf{r}$. The perturbation \textbf{$\xi$}=$(\xi _x,\xi _y, \xi _z)$ in the position of the bound electron due to the random zero-point vacuum field $\textbf{E}_0$ causes a variation in the potential energy

\begin{equation}
V(\textbf{r}+\textbf{$\xi$})=V(\textbf{r})+\textbf{$\xi$} \cdot \textbf{$\nabla$} V(\textbf{r})+\frac{1}{2}\left({\textbf{$\xi$} \cdot \textbf{$\nabla$}}\right)^{2} V(\textbf{r})...
\end{equation}
Because of the harmonic time dependence of the vacuum field, $\la \textbf{$\xi$} \ra$ vanishes and the radiative shift is given approximately by the vacuum expectation value of the last term: 
\begin{equation}
\Delta E^{Welton}_n=\frac{\la\textbf{$\xi$}^2 \ra }{6}\left\langle\nabla^{2} \mathrm{V}(\vec{\mathrm{r}})\right\rangle _n
\end{equation}
where we   assume  the potential has spherical symmetry,  thus $\la \xi_{x}^2 \ra =\la \xi_y^2 \ra=\la \xi_z^2 \ra= \la {\textbf{$\xi$}}^2/{3}\ra$.   Eq. 15 gives $\Delta E_n^{Welton}$ as the product of two factors, the first depending on the nature of the fluctuations
in the position of the bound electron due to the vacuum field and the second depending on the structure of the system. \textbf{$\xi$} 
is determined by $m $\textbf{$\ddot{\xi}$}=$e\textbf{E}_{0}$.  With a Fourier decomposition of \textbf{$E_{0}$} and \textbf{$\xi$}, and integrating over the frequency distribution of the vacuum field, we obtain the vacuum expectation value\cite{mil}\cite{gjmradiative}
\be
\langle(\vec{\xi})^2\rangle=\frac{2\alpha}{\pi}(\frac{\hbar}{mc})^2\int_0^{mc^2}\frac{dE}{E}.
\ee
Using the results in Eqs. 10 and 11 we can evaluate the Laplacian in Eq. 15 and obtain a shift for S states equal to \cite{mil}:
\begin{equation}
  \Delta E_{n}^{Welton} =  \frac{4mc^{2}}{3\pi}\alpha (Z\alpha)^{4}\frac{1}{n^3} \int _0^{mc^2} \frac{dE}{E}.
\end{equation}
Eq. 17 shows that the spectral density for the Welton approach is proportional to 1/E.  For the upper limit of integration, we take $mc^2$ as Bethe did. The lower limit of 0 gives a divergent shift. Sometimes a lower limit of the ground state energy is taken. On the other hand, if we happen to compare Eq. 17 to Eq. 12, we see that if we take for the lower limit the Bethe log Eq. 13, we get exactly the same total S state shift as in the approximate Bethe formalism Eq. 12.  With these limits, the RMS amplitude of oscillation of the electron bound in the Coulomb potential $\sqrt{\langle(\vec{\xi})^2\rangle}$ is about 72 fermis, about 1/740 of the mean radius of the 1S electron orbit.

Feynman proposed another approach for computing the Lamb shift based on a fundamental observation about the interaction of matter and the vacuum field\cite{feyn}. He considered a large box containing a low density of atoms in the quantum vacuum. The atoms cause a change in the index of refraction, which leads to changes in the frequencies of the vacuum field. The wavelengths remain the same. He maintained that the change in the energy of the zero point vacuum field in the box due to the frequency changes resulting from a weak perturbing background of atoms acting as a refracting medium would correspond to the self energy of the atoms, which is precisely the Lamb shift. 

Power, based on the suggestion by Feynman, considered the change in vacuum energy when N hydrogen atoms are placed in a volume V, using the Kramers-Heisenberg expression for the index of refraction $n(\omega_k)$\cite{power}\cite{mil}. The H atoms cause a change in the index of refraction and therefore a change in the frequencies of the vacuum fluctuations present. The corresponding change in vacuum energy $\Delta E$ is 
\be
\Delta E= \sum_{k} \frac{1}{n(\omega_k)}\frac{1}{2}\hbar\omega_k -\frac{1}{2}\hbar \omega_k  
\ee
where the sum is over all frequencies $\omega_k$ present. 
For a dilute gas of atoms in a level n, the index of refraction is \cite{mil}
\be
n(\omega_k)=1+\frac{4\pi N}{3\hbar} \sum_m\frac{\omega_{mn} |\textbf{d}|_{mn}^2}{\omega^2_{mn}-\omega_k^2}
\ee
where $\omega_{mn}=(E_m-E_n)/\hbar$ and $\textbf{d}_{mn}=e\textbf{x}_{mn}$, the transition dipole moment. After substituting $n(\omega_k)$ into Eq. 18, we get a divergent result for the energy shift. Following Bethe's approach, Power subtracted from $\Delta E$ the energy shift for the N free electrons, which equals the shift when $\omega_{mn}  \rightarrow 0$, with no binding energy.  After making this subtraction and converting the sum over $\omega_k$ to an integral over $\omega$, and letting $NV \rightarrow 1$ the observable shift in energy is obtained\cite{mil}: 

\be
\Delta E_n^{Power} = - \frac{2}{3\pi c^3}\sum_m \omega_{mn}^{3} |\textbf{d}_{mn}|^{2} \int_0^{mc^2/\hbar}\frac{d\omega \omega}{\omega_{mn}^{2}-\omega^{2}}  .
\ee
Noting that
\be
\langle m|\frac{\textbf{p}}{m}|n\rangle =\frac{i}{\hbar}\langle m |[H,\textbf{x}]| n\rangle
=\frac{i}{\hbar}(E_m-E_n)\langle m|\textbf{x}| n\rangle
\ee
we can show
\be
|\textbf{p}_{mn}|^2=m^2\omega_{mn}^2|\textbf{x}_{mn}|^2=\frac{m^2\omega_{mn}^2}{e^2}|\textbf{d}_{mn}|^2.
\ee
This allows us to write Power's result Eq. 20 as
\be
\Delta E_n^{Power} = - \frac{2e^2}{3\pi m^2 c^3}\sum_m \omega_{mn} |\textbf{p}_{mn}|^{2} \int_0^{mc^2/\hbar}\frac{d\omega \omega}{\omega_{mn}^{2}-\omega^{2}}  .
\ee
Writing this equation in terms of $E=\hbar \omega$ instead of $\omega$ yields
\be
\Delta E_{n}^{Power}=-\frac{2\alpha}{3\pi}(\frac{1}{mc})^{2}\sum_{m}|\textbf{p}_{mn}|^{2}(E_m-E_n)\int_0^{mc^2}\frac{EdE}{(E_{m}-E_n)^2-E^2}
\ee
We will use this equation to analyze the spectral density for Power's method, showing the spectral density is different from Bethe's at low frequencies but the same at high frequencies.  When Eq. 24 is integrated with respect to E, taking the principal value, we obtain
\be
\Delta E_{n}^{Power}=\frac{2\alpha}{3\pi}(\frac{1}{mc})^{2}\sum_{m}|\textbf{p}_{mn}|^{2}(E_m-E_n) ln[ {\frac{mc^2+(E_m-E_n)}{E_m-E_n}\times \frac{mc^2-(E_m-E_n)}{E_m-E_n}]^{1/2}}.
\ee
Except for the argument in the $ln$ function, which corresponds to the upper limit of integration, this is the same as Bethe's expression Eq. 7 for the shift.  If we assume $mc^2>>E_m-E_n$, as Bethe did, then both expressions for the total shift are identical. It is clear, however, that this approximation is not valid at high energies for the second factor in the $ln$ function in Eq. 25, which may even become less than one making the $ln$ term negative. Feynman's approach highlights the changes in the vacuum field energy due to the interactions with the H atoms.  

One assumption in the computation by Power is that the index of refraction in the box containing the atoms is spatially uniform.  We will return to this assumptions and suggest a model that predicts, for a single atom, the changes in the vacuum field energy as a function of position for each spectral component of the radiative shift.

\section{Spectral Density of the Lamb Shift} 

Our goal is to develop an expression for the energy shift of a level, in terms of the generators of the group SO(4,2), that is an integral over frequency. Then the integrand will be the spectral density of the shift, and group theoretical techniques can be used to evaluate it \cite{gjmdynam}. We derive a generating function for the shifts for all levels. We first focus on the ground state 1S level as an illustration of the results. At ordinary temperatures and pressures, most atoms are in the ground state. The radiative shift for the 1S level is \cite{gjmdynam} 
\be
\Delta {E}_{1}=\frac{4 mc^2 \alpha(Z \alpha)^{4}}{3 \pi} \int_{0}^{\phi_c} {d} \phi {e}^{\phi} \sinh \phi \int_{0}^{\infty} {ds e^{s{e^{-\phi}}}} \frac{d}{ds} \frac{1}{\left(\coth \frac{s}{2}+\cosh \phi \right)^{2}}
\ee 
 where the dimensionless normalized frequency variable $\phi$ is defined as
\be
\phi=\frac{1}{2} ln[1+\frac{\hbar \omega}{|E_1|}]
\ee
where $E_1$ is the ground state energy -13.6 eV. The cutoff $\phi_c$ corresponds to $E=\hbar\omega_c=m c^2$, 511 keV corresponding to the electron mass.

The group theoretical expression for the Lamb shift Eq. 26 is directly derived from the Klein-Gordon equations of motion using a non-relativistic dipole approximation, assuming infinite proton mass, and minimal coupling with the vacuum field. Basis states of $(1/Z\alpha)$ are used since they have no scattering states and have the same quantum numbers as the usual bound energy eigenstates \cite{gjmdynam}. The level shift is obtained as the difference between the mass renormalization for a spinless meson bound in the desired state and the mass renormalization for a free meson. Second order perturbation theory is not used. Near the end of the derivation an equation which is equivalent to Bethe's result Eq. 6 for the radiative shift can be derived by inserting a complete set of Schrodinger energy eigenstates. Thus we expect the fundamental results from Bethe's spectral density (with no approximations) and the group theoretical spectral density to be in agreement \cite{gjmradiative}\cite{gjmdynam}. 
For convenience an explanation of the basis states used to derive Eq. 26 is given in Appendix A, and the derivation of Eq. 26 is given in Appendix B since the derivation in \cite{gjmdynam} is spread in steps throughout the paper as the group theory methods are developed.

We can write Eq. 26 as an integral over $E=\hbar\omega$, which is the energy of the vacuum field in eV, and evaluate the definite integral over $s$ analytically for different values of $E$. We measure the ground state Lamb shift $\Delta E_{1}$ in eV so the spectral density of the shift $d \Delta E_{1}/dE$ is measured in eV/eV which is dimensionless:
 \be
\Delta E_1 = \int_0^{mc^2} \frac{d\Delta E_1}{dE}dE
 \ee
 where the ground state spectral density from Eq. 26 is
 \be
\frac{d\Delta E_1}{dE}=\frac{4 \alpha^{3}}{3 \pi}  {e}^{-2\phi} \sinh \phi \int_{0}^{\infty} {ds e^{s{e^{-\phi}}}}\frac{1} {\sinh(\frac{s}{2})^{2}} \frac{1}{\left(\coth \frac{s}{2}+\cosh \phi \right)^{3}}.
\ee

Fig. 1 shows a logarithmic plot (ordinate is a log, abscissa is linear) of the spectral density $\frac{d\Delta E_1}{dE} $ of the ground state Lamb shift with Z=1 over the entire range of energy $E$ computed from Eq. 29 using Mathematica. The spectral density is largest at the lowest energies, and decreases monotonically by about 4 orders of magnitude as the energy increases to 511 eV. The ground state shift is the integral of the spectral density from energy 0 to 511 keV.  Fig. 2 is a loglog plot (both ordinate and abscissa are log) of the same information.  The use of the loglog plot expands the energy range for each decade, revealing that for energy above about 1000 eV the slope is approximately -1, indicating that the spectral density is nearly proportional to $1/E$. For energy below about 10 eV, the spectral density in Fig. 2 is almost flat, corresponding to a linear decrease as energy increases, with a maximum spectral density at the lowest energy computed, as shown in Fig. 3. 

Fig. 2 shows that there are essentially two different behaviors of the spectral density. For values of the energy E of the vacuum field that are about 10 eV and below, in the range of the changes in energy for bound state transitions, the spectral density corresponds to the near horizontal portion of the spectral density in Fig. 2, and when E is much larger than the bound state energies, the spectral density goes as 1/E. 
 
 Fig. 3 shows linear plots (linear in ordinate and abscissa) of the spectral density of the shift for the ground state computed from Eq. 29 for several lower energy regions. Fig 3a shows a linear decrease in the spectral density as the energy increases over the small energy interval plotted. Fig 3b show a linear decrease of about 15\% as the energy increases from 0 eV to 3 eV.  Fig. 3c shows that the spectral density decreases by a factor of about 4 as the energy increases from 0 eV to 100 eV.  In the low frequency limit, the spectral density decreases linearly from the asymptotic constant value as the energy increases. 

\begin{figure}
\centering
    \includegraphics[width=0.9\linewidth, height=5cm]{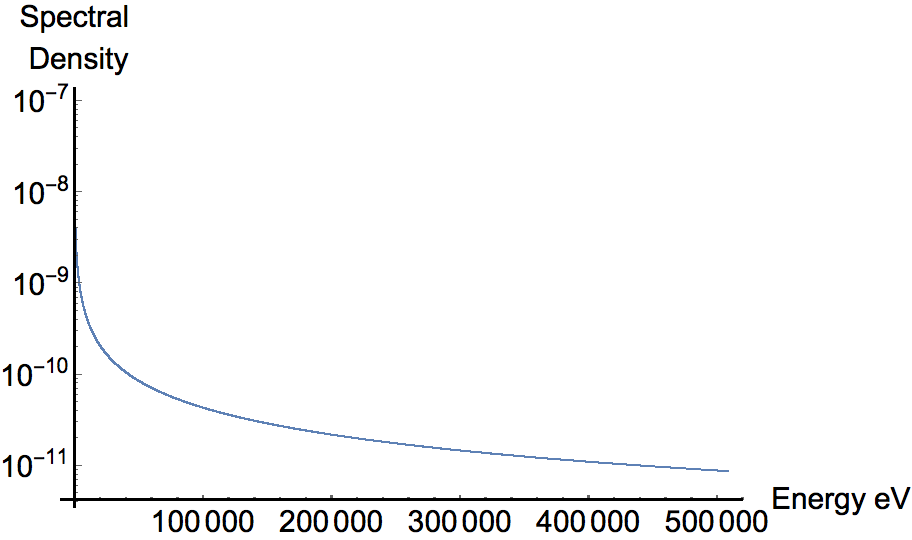}
    \caption{Plot of the log of the spectral density of the ground state Lamb shift from the group theoretical expression Eq. 29 on the vertical axis versus the energy in eV from 0 to 510 keV on the horizontal axis.}
    \label{Fig. 1}
\end{figure} 

 \begin{figure}[h]
    \centering
    \includegraphics[width=0.9\linewidth, height=5cm]{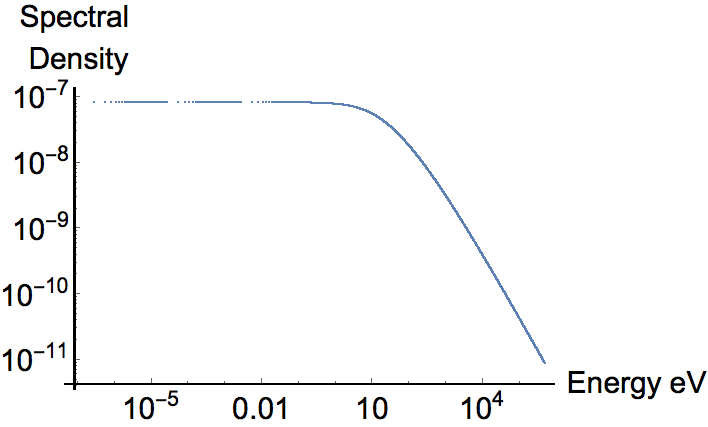}
    \caption{This loglog plot shows the log of the spectral density of the ground state shift from the group theoretical expression Eq. 29 on the vertical axis versus the log of the energy in eV. From about 0 eV to 10 eV, there is a slow linear decrease in the spectral density. For energies above about 100 eV, the behavior is dominated by a 1/energy dependence.  }
    \label{Fig. 2}
\end{figure}
\begin{figure}[p]
\begin{subfigure}{\textwidth}
\includegraphics[width=0.95\linewidth, height=4.0cm]{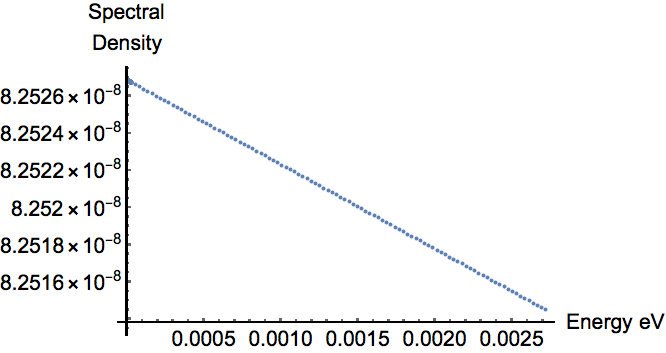} 
\caption{Linear decrease in ground state spectral density at very low energies. Note ordinate changes very little over small energy region plotted. }
\label{fig3a}
\end{subfigure}
\begin{subfigure}{\textwidth}
\vspace{10pt}
\includegraphics[width=0.95\linewidth, height=4.0cm]{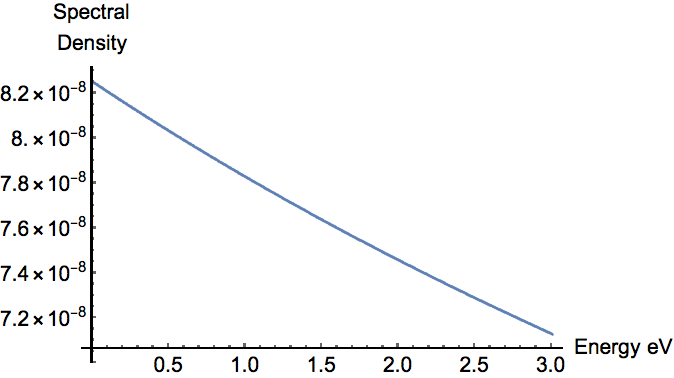}
\caption{Near linear change in ground state spectral density for visible and near IR energies. The contribution to the total shift for energies below 3 eV is about 0.7\% .}\
\label{fig3b}
\end{subfigure}
\vspace{10pt}
\begin{subfigure}{\textwidth}
\vspace{5pt}
\includegraphics[width=0.95\linewidth, height=4.0cm]{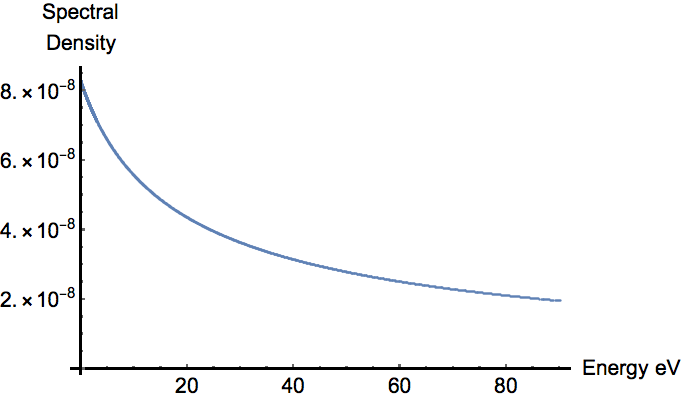}
\caption{Ground state spectral density calculated for energies below 80 eV, which contribute about $8.6\%$ to the total shift.}\
\label{fig3c}
\end{subfigure}
\caption{Linear plot of the ground state spectral density as a function of eV calculated from group theory, plotted as a function of energy for low and mid energies. From about 0 eV to 10 eV, the spectral density decreases linearly from its maximum value at the origin which corresponds to 0 eV for all graphs.}
\label{fig3}
\end{figure} 
From explicit evaluations, we will show in Section 4 that for shifts in S states with principal quantum number n, the asymptotic spectral density for large $E$ is proportional to $\alpha(Z\alpha)^4 (1/n^3)$, and show in Section 5 that as the energy E goes to zero, the spectral density increases linearly, reaching a maximum value that is proportional to $\alpha(Z\alpha)^2(1/n^2)$.  An approximate fit to the ground state data in Fig. 1 is
\be
\frac{d\Delta E_1^{Fit}}{dE}=A\frac{(1+e^{-B E})}{(E+C)}.
\ee
 where $A=4.4008\times10^{-6}$, $B=0.08445$, $C=106.79$. The fit is quite good at the asymptotes and within 10\% over the entire energy range.

We can use the spectral density shown in Fig. 1 or 2 in order to determine the contribution to the total ground state shift from different energy regions.  If we integrate the spectral density from 0 eV to energy $E$, we obtain the value of the partial shift $\Delta_1(E)$ that these energies (0 eV to $E$ eV) contribute to the total shift $\Delta E_1$ for the ground state.  In Fig. 4 we have plotted $\Delta_1(E)/\Delta E_1$, which is the fraction of the total shift $\Delta E_1$ due to the contributions from energies below $E$, as a function of $E$.  Fig. 4a shows that almost 80\% of the shift comes from energies below about 100,000 eV.  Fig. 4b shows that about half the total shift is from energies below 9050 eV. Fig. 4c shows that energies below 100 eV contribute about 10\% of the total shift. Energies below 13.6 eV contribute about 2.5\% while energies below 1 eV contribute about 1/4\% of the total. As Fig. 4c shows, the fraction of the total shift increases linearly for E<10 eV, corresponding to the nearly horizontal portion of the shift density for E<10 eV, as shown in Fig. 2. The contribution to the total 1S shift for the visible spectral interval 400-700 nm (1.770 eV to 3.10 eV) is about $1.00342 \times  10^{-7}$ eV or about 3/10 \% of the total shift.

The relative contribution to the total shift per eV is much greater for lower energies. For example, half the 1S shift corresponds to energies 0 to 9000 eV, but only about 0.2\% corresponds to 500,000 to 509,000 eV. The largest contribution to the shift per eV is at the lowest energies, which have the steepest slope of the spectral density curve in Fig. 1, about 1000 times greater than the slope for the largest values of the energy. But the total range for the large energies, from 9050 to 510,000 is so large that the absolute contribution to the total shift for large energies is significant. 

\begin{figure}[p]
\begin{subfigure}{\textwidth}
\includegraphics[width=0.95\linewidth, height=4.2cm]{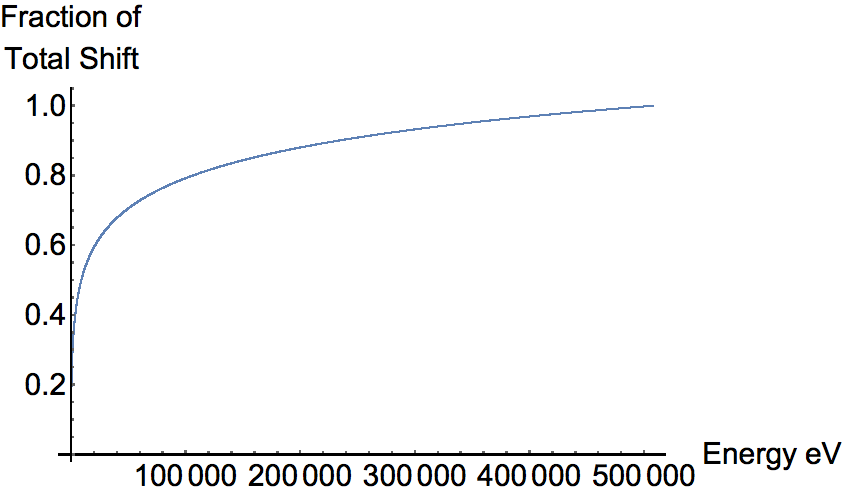} 
\caption{Fraction of the 1S shift due to energies from 0 to $E$ plotted versus $E$ on the abscissa, for $0<E<510$ keV. }
\label{fig4a}
\end{subfigure}
\begin{subfigure}{\textwidth}
\vspace{10pt}
\includegraphics[width=0.95\linewidth, height=4.0cm]{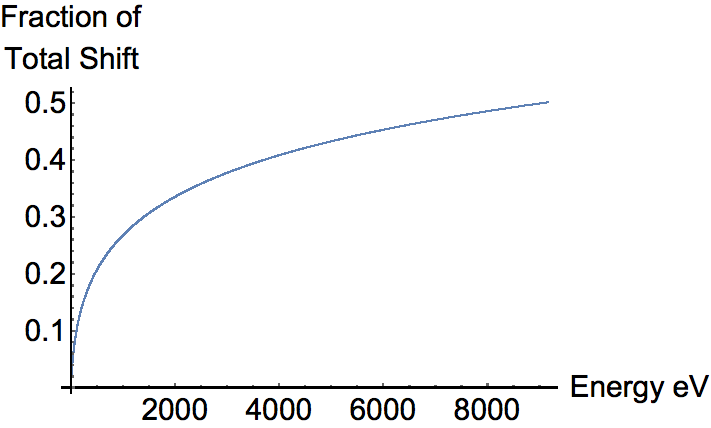}
\caption{Fraction of the 1S shift due to energies below $E$ plotted versus $E$, for $0<E<9000$ eV.}\
\label{fig4b}
\end{subfigure}
\vspace{10pt}
\begin{subfigure}{\textwidth}
\vspace{5pt}
\includegraphics[width=0.95\linewidth, height=4.0cm]{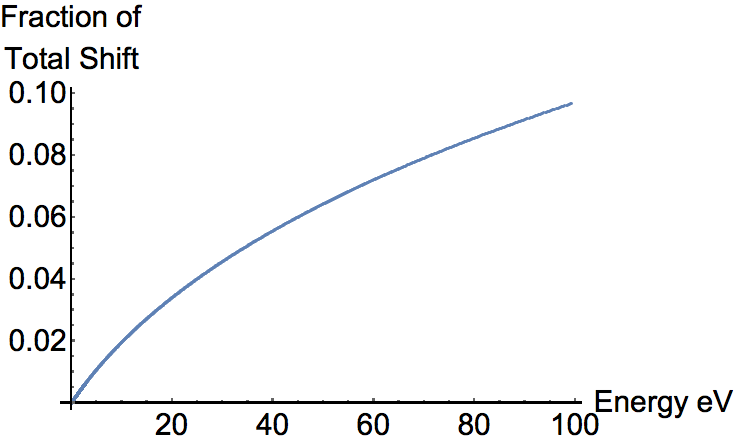}
\caption{Fraction of the 1S shift due to energies from 0 to $E$ plotted versus $E$ on the abscissa, for $0<E<100$ eV. Energies below 30 eV account for about 0.05 of the total shift. The variation is linear for E<10 eV. }\
\label{fig4c}
\end{subfigure}
\caption{The ordinate is the fraction of the ground state shift $\Delta E_1$ due to vacuum field energies between 0 and E, plotted as a function of E on the abscissa. This plot is obtained by integration of the spectral density from Eq. 29, shown in Fig. 1. The plot is linear in the ordinate and abscissa. The origin corresponds to (0,0) for all plots.}
\label{fig4}
\end{figure}

For the ground state Fig. 5 shows how the dominant terms for different $m$ in the Bethe sum over states in Eq. 6 contribute to the full spectral density obtained from group theory Eq. 29. Each such term in the Bethe sum could be interpreted as corresponding to the shift resulting from virtual transitions from state $n$ to state $m$ occurring due to the vacuum field.  Each term shown has a behavior similar to that of the full spectral density, but the magnitudes decrease as the transition probabilities decrease.
\begin{figure}[h]
    \centering
    \includegraphics[width=0.9\linewidth, height=5cm]{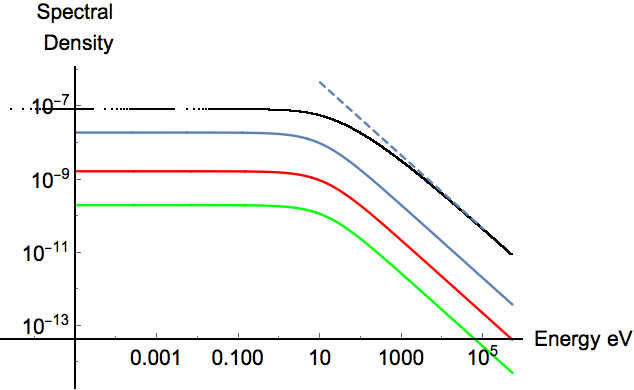}
    \caption{This loglog plot shows the 1S spectral density from group theory Eq. 29 in black, and the contributions to this shift in the Bethe formalism for the transition $1S\rightarrow 2P$ (blue), $1S\rightarrow 4P$ (red), $1S \rightarrow 8P$ (green).  The dashed blue line shows the high frequency $1/E$ asymptote. The black line is the complete spectral density which is the summation of the contributions from all transitions.}
    \label{fig 5}
\end{figure}

Fig. 6 shows the spectral densities for 1S (black) and 2S (orange) shifts.  The shapes are similar but the spectral density for the 1S shift is about eight times as large at high frequencies and about four times as large at low frequencies, factors that we will derive explicitly by considering the asymptotic forms of the spectra density for S states with different principal quantum numbers.  Both have a $1/E$ high frequency behavior.  The s integration in the group theoretical calculation for the 2S state diverges for energies below 10.2 eV due to a non-relativistic approximation, but the spectral density of the shift can be obtained from a low energy approximation, Eq. 47, to the group theory result, which we derive in Section 5.  

\begin{figure}[h]
    \centering
    \includegraphics[width=0.9\linewidth, height=5cm]{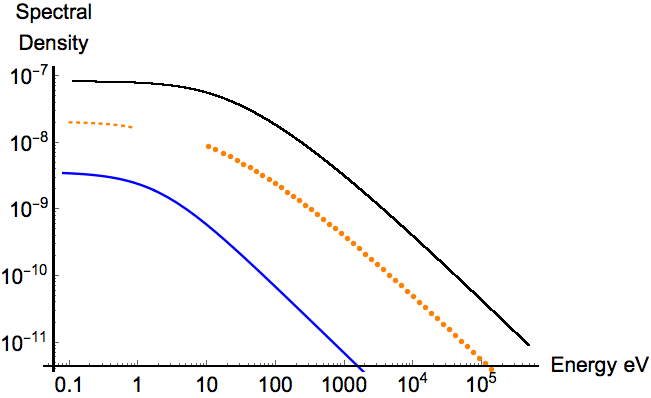}
    \caption{This loglog plot shows the log of the  group theoretical spectral density for the 1S (black) and 2S (orange) shifts on the vertical axis versus the log of the frequency in eV. The dashed orange curve below 1 eV is a 2S low energy approximation Eq. 47 from group theory or the Bethe formula. The blue is the largest single contribution in the Bethe formalism to the 2S shift for the transition $2S\rightarrow 3P$.}
    \label{fig 6}
\end{figure}

\begin{figure}[h]
    \centering
    \includegraphics[width=0.9\linewidth, height=5cm]{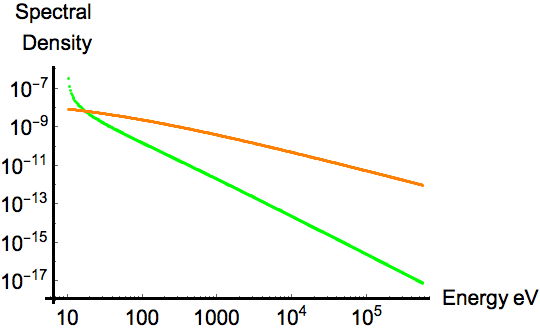}
    \caption{This loglog plot shows the log of the absolute value of the spectral density on the vertical axis versus the log of the frequency in eV for the 2S shift (orange), which goes as $1/E$ for large $E$,  and for the 2P shift (green), which goes as $1/E^2$ for large $E$. At 511 keV, the 2P spectral density is about 5 orders of magnitude smaller than the 2S spectral density.  Below 20 eV, the absolute value of the 2P spectral density is greater than the 2S spectral density. Note that the 2P spectral density is actually negative and the 2S spectral density is positive.}
    \label{fig 7}
\end{figure}

\begin{figure}[h]
    \centering
    \includegraphics[width=0.9\linewidth, height=5cm]{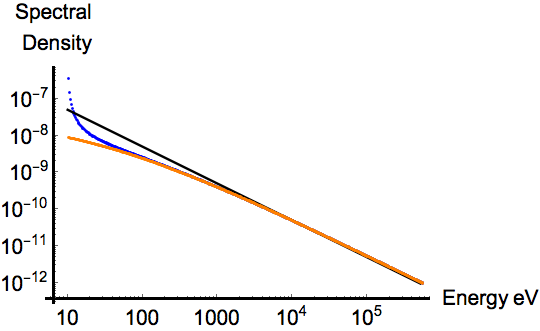}
    \caption{This loglog plot shows the log of the spectral density for the 2S shift (orange) and the 2S-2P Lamb shift (blue) versus the log of the energy. The solid black line is the $1/E$ asymptote.}
    \label{fig 8}
\end{figure}

We can define the spectral density $\frac{d\Delta E_n}{dE}$ for a state $n$ in a convenient form suggested by Eq. 29,
\be
\frac{d \Delta E_n}{dE}=\frac{4 \alpha^{3}}{3 \pi} \int_0^{\infty}ds W_n(s,\phi_n) \hspace{20pt}\text{where}\hspace{10pt}\phi_n=\ln{[1+\frac{E}{|E_n|}]}
\ee
where the energy for state $n$ is $E_n=-mc^2(Z\alpha)^2/2n^2$. 
From our group theoretical results, we have for the 2S-2P Lamb shift \cite{gjmdynam}
 \be
 W_{2S-2P}(s,\phi_2)=\frac{4 e^{(2s e^{-\phi_2}+\phi_2)} \sinh ^3(\phi_2) \text{csch}^2\left(\frac{s}{2}\right)}{\left(\cosh (\phi_2)+\coth \left(\frac{s}{2}\right)\right)^5}
 \ee
 and for the 2P shift \cite{gjmdynam}:
 \be
W_{2P}(s,\phi_2)=-\frac{e^{(2s e^{-\phi_2} +\phi_2)} \sinh (\phi_2) \text{csch}^4\left(\frac{s}{2}\right) (\cosh (\phi_2) \sinh (s)+\cosh (s)-3)}{2 \left(\cosh (\phi_2)+\coth \left(\frac{s}{2}\right)\right)^5}
\ee
The spectral density of the 2P shift has a very different behavior from the spectral density of the 2S shift (Fig. 7). It is negative and and it falls off as $1/E^{2}$. The shift is negative because the dominant contribution to the shift is from virtual transitions from the 2P state to the lower 1S state, with an energy difference of about 10.2 eV. For frequencies below about 20 eV, the absolute value of the spectral density of the 2P shift increases rapidly in magnitude as the energy is reduced and is much bigger than the spectral density for the 2S shift. The 2S shift cannot have a negative contribution from the lower 1S state since the transition 2S->1S is forbidden by the conservation of angular momentum. The classic Lamb shift arises from the difference between the two spectral densities, so the negative 2P spectral density actually increases the 2S-2P Lamb shift as the energy decreases (Fig. 8). The total 2P shift is about 0.3\% percent of the 2S shift. Bethe also computed a negative contribution for the shift from the 2P state\cite{bandsbook}.   

\subsection{Comparing the Ground State Group Theoretical Lamb Shift Calculations
to Those of Bethe, Welton, and Feynman}  

Integrating the group theoretical spectral density Eq. 29 from near zero energy $(5.4 x 10^{-7}$ eV) to 511 keV, about the rest mass energy of the electron, gives the 1S shift of $3.4027 x 10^-5$ eV, in agreement with the numerical result of Bethe and Salpeter summing over states and using the Bethe log approximation,  $3.392 x 10^-5$ eV, to about $0.3\%$ \cite{bethe}. 

 Bethe and Salpeter reported that the ground state Bethe log Eq. 13, which is a logarithmically weighted average value of the excitation of the energy levels contributing to the radiative 1S shift, was 19.77 Ry or 269 eV \cite{bandsbook}. Because of the weighting, it is not clear how one should interpret this value, other than it indicates that high energy photons and scattering states contribute significantly to the shift.  As we have noted, our group theoretical method does not provide an equivalent weighted average value for direct comparison.  

Although the methods of Bethe, Welton, and Power as defined all give approximately the same value for the 1S shift, which equals the integral of the spectral density in our approach, they differ significantly in their frequency dependence, which we will now examine.

\section{The  
Spectral Density of The Lamb shift 
at High Frequency}
The form for $d \Delta E_n/d E$, which is the Lamb shift spectral density for level $n$, can be obtained at high energies from 1) the classic calculation by Bethe using second order perturbation theory; 2) the calculation by Welton of the Lamb shift; 3) the calculation of Power of the Lamb shift based on Feynman's approach; and 4) our group theoretical calculation. 

The spectral density for level $n$ can be written from Bethe's expression Eq. 6   
\be
\frac{\Delta E_{n}^{Bethe}}{\Delta E}=\frac{2\alpha}{3\pi}(\frac{1}{mc})^{2}\sum_{m}|\textbf{p}_{mn}|^{2}(E_n-E_m)\frac{1}{E_n-E_m-E}   .
\ee
 If we are evaluating the spectral density for the ground state $n=1$, $Z=1$, then $E_1=-13.613$eV, and for the bound states $E_m= -13.613 eV/m^2$. For scattering states $E_m$ is positive. Hence the denominator is negative for all terms in the sum over $m$ and never vanishes, and the spectral density is positive, and the ground state shift is positive as it must be. For large values of $E$, we can make the approximation 
\be
\frac{\Delta E_{n}^{Bethe}}{\Delta E}|_{E-> \infty} =\frac{2\alpha}{3\pi}(\frac{1}{mc})^{2}\sum_{m}|\textbf{p}_{mn}|^{2}(E_m-E_n)\frac{1}{E} .
\ee
The summation can be evaluated using the dipole sum rule Eq. 9, and Eqs. 10 and 11 for the Coulomb S state wavefunction, obtaining the final result for the high frequency spectral density for S states with principal quantum number $n$
\begin{equation}
 \frac{d\Delta E_{n}^{Bethe}}{dE}|_{E->\infty} =  \frac{4mc^{2}}{3\pi}\alpha (Z\alpha)^{4}\frac{1}{n^3}\frac{1}{E}  .
\end{equation}
The result highlights the $1/E$ divergence at high frequencies, and shows the presence of a coefficient proportional to $1/n{^3}$. To put a scale on the coefficient, we note that the high frequency spectral density can be written as $(8/3\pi)(\alpha(Z\alpha)^2/n)(E_n/E)$.

The spectral density for all frequencies from Welton's model, Eq. 17 , is identical to this high frequency limit of Bethe's calculation. Thus at low frequencies, the spectral density for Welton's calculation diverges as $1/E.$ Because of the expectation value of the Laplacian, Welton's approach predicts a shift only for S states.  Its appeal is that it gives a clear physical picture of the primary role of vacuum fluctuations in the Lamb shift and shows the presence of the $1/E$ characteristic behavior. To obtain a level shift, it requires providing a low energy limit for the integration. As we have noted, if the lower limit is the Bethe's log average excitation energy, 269 eV for n=1, and the upper limit $mc^2$ then Welton's total 1S shift agrees with Bethe's. A choice of this type works since 1) it does not include any contributions from energies below 269 eV and 2) it gives a compensating contribution for energies from 269 eV to about 1000 eV that is larger than the actual spectral density, as shown in Fig. 4, and 3) above about 1000 eV, Welton's model gives the same $1/E$ spectral density as Bethe.

The spectral density for Power's model can be obtained from Eq. 24

\be
\frac{\Delta E_{n}^{Power}}{dE}=-\frac{2\alpha}{3\pi}(\frac{1}{mc})^{2}\sum_{m}|\textbf{p}_{mn}|^{2}(E_m-E_n)\frac{E}{(E_{m}-E_n)^2-E^2}
\ee
Letting E become large, we see the result is identical to the high frequency limit Eq. 35 for the Bethe formalism and the Welton model so we have
\be
\frac{\Delta E_n^{Power}}{dE}|_{E->\infty} = \frac{4 m c^{2}}{3\pi}\frac{\alpha(Z\alpha)^{4}}{n^{3}}\frac{1} {E}.
\ee

Thus we find for S states a $1/E$ dependence of the high frequency spectral density, corresponding to the logarithmic divergence at high frequency.  We can write this high energy theoretical result in a form allowing easy comparison to the calculated group theoretical spectral density eV/eV:

\be
\frac{d\Delta E_n^{Bethe}}{d E}|_{E->\infty}
 = \frac{4 m c^{2}}{3\pi}\frac{\alpha(Z\alpha)^{4}}{n^{3}}\frac{1}{E}.
 \ee
The spectral density goes as $1/n^3$ for S states.  For the ground state $n=1, Z=1$ we have

\be
\frac{d\Delta E_1^{Bethe}}{d E}|_{E->\infty}
=4.488 \times 10^{-6} \frac{1}{E}
\ee
A fit to the last two data points near 510 KeV in the group theoretical calculations gives:
\be
\frac{d \Delta E_1^{GTcalc}}{dE}|_{E->\infty}
=4.4008 \times 10^{-6} \frac{1}{E}  .
\ee
The coefficients differ by about $2\%$. Fig. 9 is a plot of the ground state group theoretical calculated spectral density (red) from Eq. 29 and the theoretical high energy $1/E$ function from Bethe, Power and Welton, Eq. 40 (black), and the difference times a factor of 10. The asymptotic theoretical result agrees with the full group theoretical calculation from Eq. 29 to within about $2\%$ at 511 keV, and to about $6 \%$ at 50 KeV.  It is notable that the high frequency form is a reasonable approximation down to 50 keV.  Indeed, the Welton approach is based on this observation; it has the same $1/E$ energy dependence at all energies. 

\begin{figure}
    \centering
     \includegraphics[width=0.9\linewidth, height=5cm]{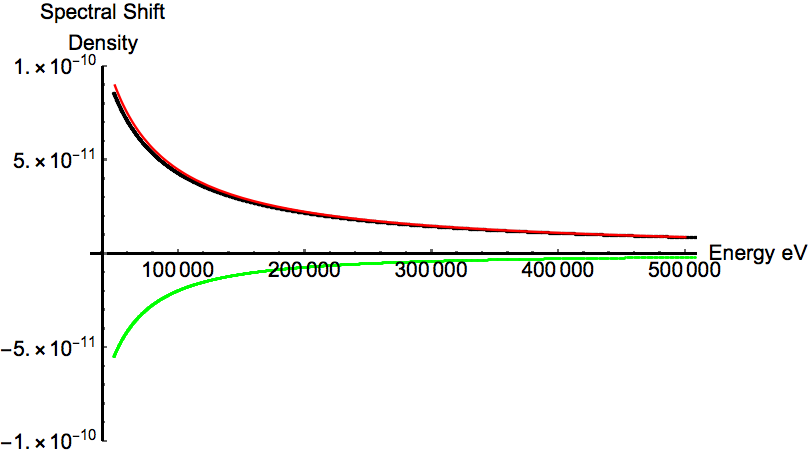}
    \caption{Top red curve is the 1S group theoretical calculated spectral density Eq. 29, slightly lower black curve is the $1/E$ asymptotic model Eq. 39, and the bottom green curve is the difference times 10, plotted for the interval 50-510keV. Both axes are linear.}
    \label{fig 9}
\end{figure}

\section{Spectral Density of the Lamb Shift at Low Frequency}

We can obtain a low frequency limit of the spectral density of the Lamb shift from the Bethe spectral density Eq. 34.  For small values of $E$, the spectral density can be expanded to first order in E, giving 

\be
\frac{\Delta E_{n}^{Bethe}}{dE}|_{E->0}=\frac{2\alpha}{3\pi}(\frac{1}{mc})^{2}\sum_{m}|\textbf{p}_{mn}|^{2}(1-\frac{E}{E_m-E_n})   .
\ee
Since the sum is over a complete set of states $m$ including scattering states we can evaluate the first term in parenthesis using the sum rule
\be
\sum_{m}|\textbf{p}_{mn}|^{2}=-2mE_n=(mc)^2\frac{(Z\alpha)^2}{n^2} .  
\ee
For the second term we use Eq. 22 and the Thomas-Reiche-Kuhn sum rule \cite{sakurai}
\be
\sum_{m}\omega_{mn}|\textbf{d}_{mn}|^{2}=\frac{3 e^2 \hbar}{2m}
\ee
to evaluate the resulting summation. The final result for $E\rightarrow0$ is
\be
\frac{\Delta E_{n}^{Bethe}}{dE}|_{E->0}=\frac{2\alpha}{3\pi}\frac{(Z\alpha)^2}{n^2}-\frac{\alpha}{\pi mc^2}E.
\ee
 The corresponding spectral density for $n=1, Z=1$ is
\be
\frac{d\Delta E_{1}^{Bethe}}{dE}|_{E->0}
=\frac{4\alpha \times 13.6}{3\pi mc^2}(1-\frac{3E}{4\times 13.6})= 8.253 \times 10^{-8}(1 - 0.0551E) .
\ee
As E decreases to zero, the spectral density increases linearly to a constant value $\frac{4\alpha}{3\pi}\frac{|E_n|}{mc^2}=2\alpha^3Z^2/3\pi n^2 =8.253\times 10^{-8}/n^{2}$. The intercept goes as $1/n^2$, but the slope $\alpha/\pi m c^2$, which has a remarkable simple form, is independent of $n$.

If we take the low frequency limit of the group theoretical result analytically, we  obtain exactly the same result as in Eq. 45 from the Bethe formulation
\be
\frac{d\Delta E_{1}^{GTheory}}{dE}|_{E->0}=\frac{d\Delta E_{1}^{Bethe}}{dE}|_{E->0}= \frac{2\alpha}{3\pi}\frac{(Z\alpha)^2}{n^2}-\frac{\alpha}{\pi mc^2}E.
\ee
Fig. 3 shows the results of group theoretical calculations of the spectral density of the ground state Lamb shift for different energy regions, showing the near linear increase in the spectral density as the frequency decreases from 80 eV to $10^{-5}$ eV. For low values of E, the slopes and intercept agree within about two tenth of a percent with the theoretical values from Eq. 47.   

To explore Power's approach at low frequency, we can let $E$ become very small in the spectral density Eq. 37, giving 
\be
\frac{\Delta E_n^{Power}}{dE}|_{E->0} = - \frac{2\alpha}{3\pi mc^3}\sum_m |\textbf{p}_{mn}|^2 \frac{E}{E_m-E_n}  
\ee
which is identical to the second term in the low E approximation to the Bethe result Eq. 42 so we have: 

\be
\frac{\Delta E_{n}^{Power}}{dE}=-\frac{1}{\pi}\frac{\alpha}{m c^2}E.
\ee
This result Eq. 49 is identical to the frequency dependent term in Eq. 47, which is the low frequency spectral density from the Bethe approach and from the group theoretical expression. However, in the low frequency limit based on Power's expression for the spectral density, the constant term that is present in the other approaches does not appear. This a consequence of the form used for the index of refraction, which assumes that real photons are present that can excite the atom with resonant transitions. More sophisticated implementations of Feynman's proposal may avoid this issue.

    


\section{Comparison of the Spectral Energy Density of the Vacuum Field and  the Spectral Density of the Radiative Shift}

The theory of Feynman proposes that the vacuum energy density in a large box containing H atoms, which we assume are all in the 1S ground state, increases uniformly with the addition of the atoms. He maintains that the total vacuum energy in the box increases by the Lamb shift times the number of atoms present. If we had one atom in a very large box, we would not expect the change in energy density to be uniform but more concentrated near the atom. To develop a model of the spatial dependence of the change in energy density for one atom, we can use the close relationship between the vacuum field and the radiative shift. The spectral densities of the ground state shift and of the quantum vacuum with no H atoms present are both know.  In the box the vacuum field density must increase so that the integral gives the 1S Lamb shift.  The spectral energy density of the vacuum field with no H atom present is equal to \cite{mil}
\be
\rho_0(\omega)=\frac{\hbar\omega^3}{2\pi^2c^3}
\ee
where c is the speed of light in cm/sec and $\omega$ is in $sec^{-1}$.  If we measure frequency in eV so $\hbar \omega=E$ then the vacuum spectral energy density in $1/cc$ is 
\be
\rho_0(E)=\frac{E^3}{2\pi^2 \hbar^3 c^3} . 
\ee
and $\int_{E_1}^{E_2}\rho_0(E)dE$ would be the energy density eV/cc in the energy interval $E_1$ to $E_2$. The question we are addressing is: what volume of vacuum energy of density $\rho_0(E)$ is required to supply the amount of energy needed for the radiative shift? We can express the total radiative shift $\Delta E_1$ as the integral of the vacuum energy density $\rho_0(E)$ over an effective volume $V_1(E)$ 
\be
\Delta E_1=\int_0^{mc^2} dE \rho_0(E) V_1(E)
\ee
where we use the same upper limit for $E$ as in all of our calculations.  Recall our definition of the spectral shift Eq. 28:
\be
\Delta E_1=\int_0^{mc^2} dE \frac{\Delta E_1}{dE}.
\ee
By comparison of Eq. 52 and Eq. 53 we determine that to insure energy balance at each energy E, the effective spectral volume $V_1(E)$ is
\be
V_1(E)=\frac{d\Delta E_1}{dE}\frac{1}{\rho_0(E)}.
\ee
The spectral volume $V_1(E)$ has the dimensions of $cc$ and contains the amount of vacuum energy at energy value $E$ that corresponds to the ground state spectral density at the same energy $E$. In Fig. 10, for the 1S ground state radiative shift, we plot the log of the spectral volume $V_1(E)$ on the y-axis in units of cubic Angstroms versus the log of the energy $E$ in eV on the x-axis. 
\begin{figure}[h]
\centering                \includegraphics[width=0.94\textwidth,height=5cm]{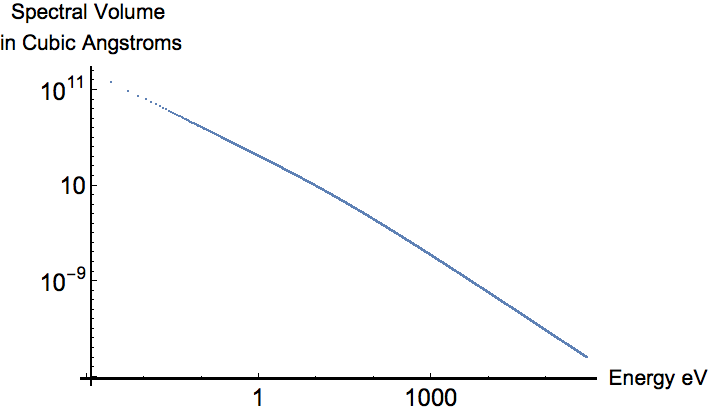}         \caption{This loglog plot shows the spectral volume $V_{1}(E)$ as a function of $E$.  The spectral volume $V_1(E)$ contains the free field vacuum energy at energy value $E$ that corresponds to the ground state shift spectral density at the same energy $E$.}
\label{fig 10}
\end{figure}
For energies above about 100 eV, the spectral volume is less than 1 cubic Angstrom, approximately the volume of the ground state wavefunction. For an energy of 1 eV, the spectral volume is $11850 A^3$, corresponding to a sphere of radius about 14 A. This calculation predicts that there is a sphere of positive vacuum energy of radius 14 A around the atom corresponding to the 1 eV shift spectral density. Fig. 11 shows the radius of the spherical spectral volume $V_1(E)$ for energies from 0.05 eV, with spectral radius of 278 A, to 23 eV, with radius 0.49 A. 
\begin{figure}
    \centering
    \includegraphics[width=0.94\textwidth,height=5cm]{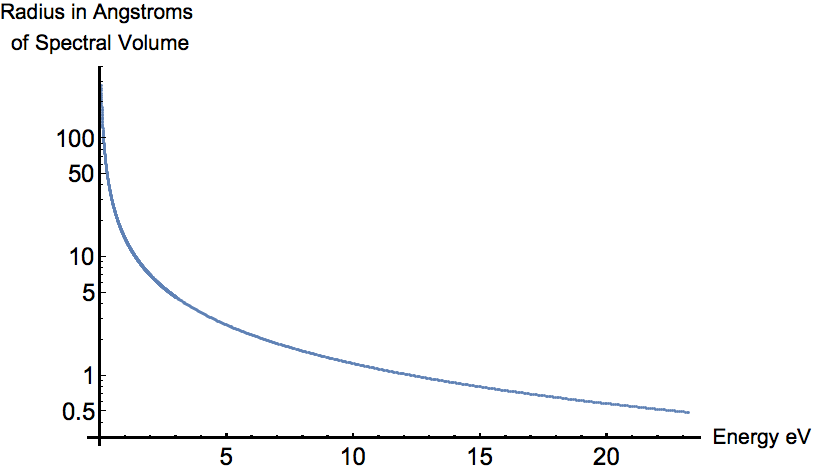}
    \caption{This plot shows the log of the radius in Angstroms of the spherical spectral volume $V_1(E)$ as a function of the vacuum field energy E from 0.05 eV to 23 eV.}
    \label{fig:my_label}
\end{figure}

\section{Conclusion}

The non-relativistic Lamb shift can be interpreted as due to the interaction of an atom with the fluctuating electromagnetic field of the quantum vacuum. We introduce the concept of a spectral shift density which is a function of frequency $\omega$ or energy $E=\hbar \omega$ of the vacuum field. The integral of the spectral density from E=0 to the rest mass energy of an electron, 511 keV, gives the radiative shift. We report on calculations of the spectral density of the level shifts for 1S, 2S and 2P states based on a group theoretical analysis and compare the results to the spectral densities implicit in previous calculations of the Lamb shift.  The group theoretical calculation provides an explicit form for the spectral density over the entire spectral range. Bethe's approach requires a summation over an infinite number of states, all bound and all scattering, to obtain a comparable spectral density. We compare all approaches for asymptotic cases, for very large and very small energies E.

The calculations of the shift spectral density provide a new perspective on radiative shifts. The group theory approach as well as the approaches of Bethe, Power, and Welton all show the same $1/E$ high frequency behavior for S states above about $E=\hbar\omega$= 1000 eV to E=511 keV, namely a spectral density for S states equal to  $(4/3\pi)(\alpha(Z\alpha)^4mc^2/n^3)(1/E)$ for states with principal quantum number $n$.  Since our group theory calculation shows that about 76\% of the ground state 1S shift is contributed by E above 1000 eV, this is essentially why all the approaches give approximately the same result for the 1S Lamb shift. 

Only the Bethe and group theory calculations have the correct low frequency behavior. We find that for S states the spectral density increases linearly as E approaches zero. Its maximum value is at E=0 and for S states equals $(2\alpha/3\pi)(Z\alpha)^2/n^2$.  This maximum value is about $1/(Z\alpha)^{2}$ or about $2\times 10^4$ larger than the high frequency spectral density at E=510 keV. Thus low energies contribute much more to the shift for a given spectral interval than the high energies. Energies below 13.6 eV contribute about 2.5 \%.  Because of the huge spectral range contributing to the shift, contributions to the shift from high energies are very important. Half the contribution to the 1S shift is from energies above 9050 eV. 

 The 2P shift has a very different spectral density from an S state: it is negative and has an asymptotic behavior that goes as $1/E^2$ rather than as $1/E$. Below about 20 eV, the absolute value of the 2P spectral density is much larger than the 2S spectral density and it dominates the 2S-2P shift spectral density, yet the total 2P shift is only about 0.3\% of the total 2S shift. 
\appendix
\counterwithin{equation}{section}
\renewcommand{\theequation}{\thesection\arabic{equation}}
\section[Eigenstates of |nlm;a> of 1/Z alpha]{:\hspace{15pt}Eigenstates $|nlm;a)$ of $1/Z\alpha$}

To obtain an equation for these basis states $|nlm;a)$ we write Schrodinger's equation for a charged non-relativistic particle with energy $E=-\frac{a^2}{2m}$\cite{gjmdynam}\cite{brow}
in a Coulomb potential
\be
\left[p^{2}+a^{2}-\frac{2 m\hbar c Z \alpha}{r}\right]|a\rangle=0.
\ee
There are solutions for $|a\rangle$ for certain critical values of the energy $E_n= -\frac{a_n^2}{2m}$ or equivalently when $a=a_n$ where $\frac{a_n}{mcZ\alpha}=\frac{1}{n}$. These are the usual energy eigenstates which we label as $|nlm;a_n \rangle$. Conversely we can let $a$ be fixed in value and let $Z\alpha$ have different values. If it has certain eigenvalues $Z\alpha_n$ then for any value of $a$ we can have another set of eigenvectors corresponding to eigenvalues $\frac{a}{mcZ\alpha_n}=\frac{1}{n}$. To demonstrate this
we start by inserting factors of $1 = \sqrt{ar}\frac{1}{\sqrt{ar}}$ in Schrodinger's equation Eq. A1 obtaining
\be
\left( \sqrt{ar}(p^2+a^2)\sqrt{ar}-2amZ\alpha\right)\frac{1}{\sqrt{ar}}|a \rangle= 0 .
\ee
We can rewrite this equation, multiplying successively from the left by $\frac{1}{\sqrt{ar}}$,$\frac{1}{p^2+a^2}$, and $\frac{1}{\sqrt{ar}}$, and then multiplying by $a^2$, and dividing by $mcZ\alpha$, multiplying by $\sqrt{n\hbar}$ obtaining 
\be
\left(\frac{a}{mcZ\alpha}-K_1(a)\right) \sqrt{\frac{n \hbar}{ar}}|a\rangle = 0
\ee

where
\be
K_{1}(a)=\frac{1}{\sqrt{a r}} \frac{2 a^{2}\hbar}{p^{2}+a^{2}} \frac{1}{\sqrt{a r}}
\ee
There are solutions to this equation for eigenvalues of $1/Z\alpha$ such that $\frac{a}{mcZ\alpha_n}=\frac{1}{n}$:
\be
\left(\frac{1}{n}-K_1(a)\right)|nlm;a) = 0
\ee
where
$$\sqrt{\frac{n \hbar}{ar}}|nlm;a\rangle=|nlm;a)$$

The $n\hbar$ in the square root insures the new states are also normalized to 1. The kernel $K_1(a)$ is bounded and Hermetian with respect to the eigenstates $|nlm; a)$ of $1/Z\alpha$, therefore these eigenstates of $1/Z\alpha$ form a complete orthonormal basis for the hydrogen atom. Because the kernel is bounded, there are no continuum states in this representation. To show they have the same quantum numbers as the usual states, we note when $a=a_n$ then the eigenstates of $K_1(a_n)$ becomes $|nlm; a_n)$ and these corresponds to the usual energy eigenstates $|nlm;a_n\rangle$.  We can change the value of $a$ in Eq. A5 to obtain these eigenstates using the dilation operator $D(\lambda)=e^{iS\lambda}$ where the dimensionless operator S, which is also a generator of transformations of SO(4,2), is  
\be
S=\frac{1}{2\hbar}(\bm{p}\cdot\bm{r} + \bm{r}\cdot\bm{p}).
\ee
When S operates on the canonical variables we obtain
$$D(\lambda)\textbf{p}D^{-1}(\lambda)=e^{-\lambda} \textbf{p}$$ 
$$D(\lambda)\textbf{r}D^{-1}(\lambda)=e^{\lambda} \textbf{r}.$$ 
Operating on $K_1(a)$ with $D(\lambda)$ we find
$$D(\lambda)K_1(a)D^{-1}(\lambda)=K_1(ae^{\lambda}).$$  
We can pick $\lambda$ as $$\lambda_n=ln(a_n/a)$$ so that $ae^{\lambda_n}=a_n$. Thus operating with $D(\lambda_n)$ on Eq. A5 we obtain
\be 
 \left(\frac{1}{n}-K_1(a_n)\right)D(\lambda_n)|nlm;a) = 0.  
\ee
This is the equation for the usual Schrodinger energy eigenstates so 
\be
D(\lambda_n )|nlm;a)=|nlm;a_n)=\sqrt{\frac{n\hbar}{a_n r}}|nlm;a_n\rangle.
\ee
Thus the usual Schrodinger energy eigenstates $|nlm;a_n\rangle$ can be expressed in terms of the eigenstates of $1/Z\alpha$ as
\be
|nlm;a_n\rangle = \sqrt{\frac{a_nr}{n\hbar}}D(\lambda_n)|nlm;a).
\ee
The relationship shows that complete basis functions $|nlm;a)$ of $1/Z\alpha$ are proportional to the ordinary bound state energy wavefunctions and therefore have the same quantum numbers as the ordinary bound states\cite{gjmdynam}\cite{brow}.  A comparable set of $1/Z\alpha$ eigenstates useful for momentum space calculations is derived in \cite{gjmdynam}. 

\section{:\hspace{15pt}     Derivation of Group Theoretical Formula for the Shift Spectral Density}
The group theoretical approach is based solely on the Schrodinger and Klein-Gordon equations of motion in the non-relativistic dipole approximation. We obtain a result \cite{gjmdynam}
\be
\Delta E_{NL}=\frac{2\alpha}{3\pi (mc)^2} \int_0^{\hbar \omega_c} dE  \la NL|p_i \frac{H-E_N}{H-(E_N-E) - i\epsilon}p_i|NL \ra .
\ee
where $E=\hbar \omega$, $H=\frac{p^2}{2m}-\frac{Z\alpha\hbar c}{r}$ and the states $|NL\ra$ are the usual H atom energy eigenstates. $\omega_C$ is a cutoff frequency for the integration that we will take as $\hbar \omega_c = mc^2$. If we insert a complete set of states in this expression we obtain Bethe's result Eq. 6, a step we avoid with the group theoretical approach. If we add and subtract $E$ from the numerator in Eq. B1, we find the real part of the shift is 
\be
\Delta E_{NL}=\frac{2\alpha}{3\pi (mc)^2}Re\int_0^{\hbar\omega_c} dE [\la NL|p^2 |NL \ra - E\Omega_{NL}]
\ee
where
\be
\Omega_{NL}=\la NL| p_i \frac{1}{H-E_N +\hbar\omega -i\epsilon} p_i |NL \ra.
\ee
We want to convert the matrix element $\Omega_{NL}$ to a matrix element of a function of SO(4,2) generators taken between a new set of basis states $|nlm;a)$, which are complete with no scattering states, where $a=\sqrt{2m|E|}$, and n,l,m have their usual meaning and values. The new basis states $|nlm;a)$ are eigenstates of $(Z\alpha)^{-1}$\cite{gjmdynam} \cite{brow}. Sometimes we simply write them as $|nlm)$ with the $a$ implicit.

We define a generator of SO(4,1) as $\Gamma_0=1/K_1(a)=(1/2)(\frac{\sqrt{r}p^2\sqrt{r}}{a}+ar)$ so
\be
(\Gamma_0 - n)|nlm;a)=0.
\ee
This is Schrodinger's equation in the language of SO(4,2).

We need to define several more generators. Since the algebra of SO(4,2) generators closes, commutators of generators must also be generators. 
To find $\Gamma_4$, we calculate $\Gamma_4=-i[S, \Gamma_0]$, obtaining
\be
\Gamma_4= \frac{1}{2\hbar}\left(\frac{\sqrt{r}p^2\sqrt{r}}{a}-ar\right)\hspace{20pt}\Gamma_0= \frac{1}{2\hbar}\left(\frac{\sqrt{r}p^2\sqrt{r}}{a}+ar\right)
\ee
where the generator $S$ is defined in Appendix A. The generators $({\Gamma_4, S, \Gamma_0})=(j_1,j_2,j_3)$ form a O(2,1) subgroup of SO(4,2) and $S=i[\Gamma_4,\Gamma_0], \Gamma_0=-i[S,\Gamma_4]$ and for our representations $\Gamma_{0}^2-\Gamma_{4}^2-S^2=\boldsymbol{L^2}=l(l+1)$. The scale change S transforms $\Gamma_0 \equiv \Gamma_0(a)$  according to the equation
\be
e^{i\lambda S}\Gamma_0(a)e^{-i\lambda S}=\Gamma_0(e^{\lambda}a)=\Gamma_0\cosh{\lambda}-\Gamma_4\sinh{\lambda}
\ee
and similarly
\be
e^{i\lambda S}\Gamma_4(a)e^{-i\lambda S}=\Gamma_4(e^{\lambda}a)=\Gamma_4\cosh{\lambda}-\Gamma_0\sinh{\lambda}.
\ee

Finally we define a three vector of generators proportional to the momentum 
\be
\Gamma_i = \frac{1}{\hbar}\sqrt{r}p_i \sqrt{r}.
\ee
The quantity $\bm{\Gamma} =(\Gamma_0,\Gamma_1,\Gamma_2,\Gamma_3,\Gamma_4)$ is a five vector of generators under transformations generated by SO(4,2).
For the representation of SO(4,2) based on the states $|nlm)$, all generators are Hermetian, and  $\bm{\Gamma}^2=\Gamma_A \Gamma^A=-\Gamma_0^2+\Gamma_1^2+\Gamma_2^2+\Gamma_3^2+\Gamma_4^2=1$ for our representation, and $g_{AB}=(-1,1,1,1,1)$ for $A,B=0,1,2,3,4$.  The commutators of the components of the five vector are also generators of $SO(4,2)$ transformations.

Inserting factors of $1=\sqrt{ar}\frac{1}{\sqrt{ar}}$ and using the definitions of the generators we can transform Eq. B3 to
\be
\Omega_{NL}=\frac{m\nu}{N^2}(NL|\Gamma_i\frac{}{\Gamma n(\xi) - \nu}\Gamma_i|NL)
\ee
where
\be
n^0(\xi)=\frac{2+\xi}{2\sqrt{1+\xi}}=\cosh \phi \hspace{20pt}n^i = 0 \hspace{20pt} n^4(\xi)= -\frac{\xi}{2\sqrt{1+\xi}}=-\sinh \phi  \ee
and 
\be
\xi=\frac{\hbar \omega}{|E_N|} \hspace{20pt}\nu=\frac{N}{\sqrt{1+\xi}}=Ne^{-\phi}.
\ee
From the definitions we see $\phi = \frac{1}{2}ln(1 + \xi) > 0$ and $n_A(\xi)n^A(\xi) = -1$. The~contraction over $i$ in $\Omega_{NL}$ may be evaluated using the group theoretical formula \cite{gjmdynam}:
\be 
 \sum_{B} \Gamma_Bf(n\Gamma) \Gamma^B = \frac{1}{2}(n\Gamma +1)^2 f(n\Gamma +1) + \frac{1}{2}(n\Gamma -1)^2 f(n\Gamma -1) - (n\Gamma)^2 f(n\Gamma)   .
\ee 
We apply the contraction formula to the the integral representation
\be
f(n\Gamma)=\frac{1}{\Gamma n - \nu}  =\int_0^\infty ds e^{\nu s} e^{-n\Gamma  \hspace{1pt}s}
\ee
and obtain the result
\be
\Gamma_A \frac{1}{\Gamma n - \nu} \Gamma^A  =-2 \nu \int_0^\infty ds\hspace{2pt} e^{\nu s}\frac{d}{ds}(\sinh^2 \frac{s}{2} \hspace{2pt}e^{-n\Gamma \hspace{1pt} s}).
\ee
Applying this to our expression Eq. B9 for $\Omega_{NL}$ gives
\be
\begin{aligned} \Omega_{N L} &=-2 \frac{m \nu^{2}}{N^{2}} \int_{0}^{\infty} d s e^{\nu s} \frac{d}{d s}\left(\sinh ^{2} \frac{s}{2} M_{N L}(s)\right) \\  & \vspace{5pt} -m \frac{\nu}{N^{2}} (N L|\Gamma_{4} \frac{1}{\Gamma n(\xi)-\nu} \Gamma_{4} | N L) + m\frac{\nu}{N^2} (N L| \Gamma_0 \frac{1}{\Gamma n(\xi) - \nu} \Gamma_0|NL) \end{aligned}
\ee
where
\be
M_{NL}(s) = (NL| e^{-\Gamma n(\xi)\hspace{1pt} s} |NL)  .
\ee
In order to evaluate the last two terms in Eq. B15 we use $\Gamma_0|NL)=N|NL)$ and express the action of $\Gamma_4$ on our states as $\Gamma_4=N-(1/\sinh{\phi})(\Gamma n(\xi) - \nu)$. This expression for $\Gamma_4$ is derived from Eqs. B9 and B10:  $\Gamma n(\xi) - \nu= \Gamma_0 \cosh{\phi}-\Gamma_4 \sinh{\phi}-\nu$, and then substituting Eq.B11, $\nu=N e^{-\phi}$. Using the virial theorem $(NLM|p^2|NLM)=a_N^2$, we find that the term in $p^2$ in Eq. B15 exactly cancels the last two
terms in $\Omega_{NL}$, yielding the result for the level shift
\be
\Delta E_{NL}=\frac{4 mc^2 \alpha(Z \alpha)^{4}}{3 \pi N^{4}} \int_{0}^{\phi _c} d \phi  \sinh \phi e^{\phi} \int_{0}^{\infty} d s\hspace{2pt} e^{\nu s} \frac{d}{d s}\left(\sinh ^{2} \frac{s}{2} M_{N L}(s)\right)
\ee
where
\be
\phi_c= \frac{1}{2}ln \left(1 + \frac{\hbar \omega_c}{|E_N|} \right)=\frac{1}{2}ln \left( 1+\frac{2N^2}{(Z\alpha)^2} \right).
\ee

We can derive a generating function for the shifts for any eigenstate characterized by $N$ and $L$ if we multiply Eq.B17 by $N^4 e^{-\beta  N}$ and sum
over all $N, N \ge L + 1$. To~simplify the right side of the resulting equation, we use the definition Eq. B16 and the fact that
$\Gamma_4$, ~$S$, and $\Gamma_0$ form an O(2,1) algebra so we have:
\be
\sum_{N={L}+1}^{\infty} e^{-\beta {N}} M_{NL}=\sum_{{N}={L}+1}^{\infty}( {NL} | e^{-\bm{j} \cdot \bm{\psi}}| {NL} ),
\ee
where
\be
e^{\hspace{1pt}-\bm{j}\cdot \bm{\psi}} \equiv e^{-\beta \Gamma_0} e^{-s \Gamma n(\xi)}  .
\ee
We perform a $\bm{j}$ transformation generated by $e^{i\phi S}$, such that
$
e^{-\bm{j} \cdot \psi} \rightarrow e^{-j_3 \psi}=e^{-\Gamma_0 \psi}$.
The trace is invariant with respect to this transformation so we have 
\be
\sum_{N={L}+1}^{\infty} e^{-\beta {N}} M_{NL}=\sum_{{N}={L}+1}^{\infty}( {NL} | e^{-j_3 \psi}| {NL} ) =\sum_{N={L}+1}^{\infty} e^{-N \psi}=\frac{e^{-\psi (L+1)}}{1-e^{-\psi}}  ,
\ee
where we have used $(NL|\Gamma_0)|NL)=N$.

In order to find a particular $M_{NL}$, we must expand the right hand side
of the equation in powers of $e^{-\beta}$ and equate the coefficients to those on the left hand side. Using the isomorphism between $\bm{j}$ and the Pauli $\bm{\sigma}$ matrices $(\Gamma_4, S, \Gamma_0) \rightarrow (j_1,j_2,j_3) \rightarrow (\frac{i}{2}\sigma_1,\hspace{2pt} \frac{i}{2}\sigma_2,\hspace{2pt} \frac{1}{2} \sigma_3)$
gives the result
\be
\cosh{\frac{\psi}{2}}=\cosh{\frac{\beta}{2}}\cosh{\frac{s}{2}}+\sinh{\frac{\beta}{2}}\sinh{\frac{s}{2}}\cosh{\phi}.
\ee
Rewriting this equation gives
\be
e^{+\frac{1}{2} \psi}=d e^{\frac{1}{2} \beta}+b e^{-\frac{1}{2} \beta}-e^{-\frac{1}{2} \psi}
\ee
where
\be
\begin{array}{l}d=\cosh \frac{s}{2}+\sinh \frac{s}{2} \cosh \phi \\ b=\cosh \frac{s}{2}-\sinh \frac{s}{2} \cosh \phi\end{array}   .
\ee
Let $\beta$ become very large, which implies large $\psi$, and iterate the equation for $e^{-\frac{1}{2}\psi}
$ to obtain the result
\be
{e}^{-\psi}={Ae}^{-\beta}\left[1+{A}_{1} {e}^{-\beta}+{A}_{2} {e}^{-2 \beta}+\ldots\right]   
\ee
where
$ A=1/d^{2}$ and $A_{1} =-(2/d)(b-d^{-1})$.  To obtain $M_{NL}$, we expand the right side of Eq. B21 in powers of $\psi$ 
\be
\frac{e^{-\psi (L+1)}}{1-e^{-\psi}}=\sum_{m=1}^\infty e^{-\psi(m+L)}.
\ee
For large $\beta$ it follows from Eqs. B21, B25, and B26 that
\be
\sum_{N=L+1}^\infty e^{-\beta N} M_{NL} =\sum_{m=1}^\infty \left[e^{-\beta} A(1+A_1e^{-\beta} + A_2 e^{-2\beta} +...\right]^{m+L}.
\ee
Using the multinomial theorem \cite{morse} the right side of Eq. B27 becomes
\be
\sum_{m=1}^{\infty}A^{m+L}\sum_{r,s,t,...}\frac{(m+L)!}{r!s!t!...}A_1^sA_2^t...e^{-\beta(m+L+s+2t+...)}
\ee
where $r+s+t+...=m+L$.
To obtain the expression for $M_{NL}$, we note $N$ is the coefficient of $\beta$ so $N=m+L+s+2t+...=r+2s+3t+...$  Accordingly we find
\be
M_{NL}=\sum_{r,s,t,}A^{(r+s+t+...)}\frac{(r+s+t+...)!}{r!s!t!}A_1^sA_2^t...
\ee
where r+s+t+...=N and r+s+t+...>L.  For the 1S shift, as Eq. B17 indicates, we want the matrix element $M_{10}$ which corresponds to $e^{-\beta}$ so $M_{10}=A$. For the 2S shift we have $M_{20}=A^2+AA_1$, and for the 2P shift $M_{21}=A^2$. Therefore the radiative shift for the $1S$ ground state is
\be
Re \Delta {E}_{10}=\frac{4 mc^2 \alpha(Z \alpha)^{4}}{3 \pi} \int_{0}^{\phi_c} {d} \phi {e}^{\phi} \sinh \phi \int_{0}^{\infty} {ds e^{s{e^{-\phi}}}} \frac{d}{ds} \frac{1}{\left(\coth \frac{s}{2}+\cosh \phi \right)^{2}}.
\ee
The shift for the 2S-2P level is
\be
Re(\Delta E_{20} - \Delta E_{21})= \frac{m \alpha(Z \alpha)^{4}}{6 \pi} \int_{0}^{\phi_c} d \phi e^{\phi} \sinh ^{3} \phi \int_{0}^{\infty} d s e^{2 s e^{-\phi}} \frac{d}{d s} \frac{1}{\left(\coth \frac{s}{2}+\cosh \phi\right)^{4}}.
\ee 

\section*{Funding}
This research received no external funding. The author has no conflicts of interest.
\section*{Acknowledgements} I thank Prof. Peter Milonni for many insightful and enjoyable discussions, particularly about the resonant behavior of the index of refraction and the volume of vacuum energy corresponding the the spectral density , and I thank Prof. Lowell S. Brown for his observations, especially about the 1/ frequency asymptotic behavior.


\section*{References}
\begin{thebibliography}{99}
\bibitem{astro}Choudhuri, A., \emph{Astrophysics for Physicists},Cambridge University Press, Cambridge, England, 2010.

\bibitem{espeg}d'Espagnat, B.,  \emph{Veiled Reality, An Analysis of Present-Day Quantum Mechanical Concepts}, Addison-Wesley, Reading, MA, USA, 1995

\bibitem{beyer} Beyer, A.; Maisenbacher, L.; Matveev, A.; Pohl, R.; Khabarova, K.; Grinin, A.; Lamour, T.; Yost, D.C.; Hänsch,~T.W.; Kolachevsky, N.; et al., The Rydberg constant and proton size from atomic hydrogen. \emph{Science} \textbf{2017}, \emph{358}, 79.

\bibitem{lam1}Lamb, W.; Retherford, R., Fine Structure of the Hydrogen Atom by a Microwave Method. \emph{Phys. Rev.} \textbf{1947}, \emph{72},~241.

\bibitem{bethe}Bethe, H., The Electromagnetic Shift of Energy Levels. \emph{Phys. Rev.} \textbf{1947}, \emph{72}, 339.

\bibitem{welton}Welton, T., Some observable effects of the quantum-mechanical fluctuations of the electromagnetic field. \emph{Phys.~Rev.} \textbf{1948}, \emph{74}, 1157.

\bibitem{power}Power, E.A., Zero-Point Energy and the Lamb Shift. \emph{Am. J. Phys.} \textbf{1966}, \emph{34}, 516.


\bibitem{mil}Milonni, P., \emph{The Quantum Vacuum}; Academic Press: San Diego, CA, USA, 1994.

\bibitem{passante}
Campagno, G., Passante, R., and Persico.F., \emph{Atom-field Interactions and dressed Atoms}; Cambridge University Press:Cambridge, England, 1995.

\bibitem{gjmradiative}Maclay, G. Jordan,  History and Some Aspects of the Lamb Shift, Physics 2(2), 105-149(2020).


\bibitem{feyn}Feynman, R., \emph{Solvay Institute Proceedings}:Interscience Pblishes, Inc. New York, USA, 1961.


\bibitem {lieber}Lieber, M., O(4) Symmetry of the Hydrogen Atom and the Lamb Shift, Phy. Rev 174,2037(1968). 

\bibitem{huff}Huff, R., Simplified Calculation of Lamb Shift Using Algebraic Techniques, Phys. Rev. 186,1367(1969).

\bibitem{gjmdynam}Maclay, G.Jordan, Dynamical Symmetries of the H Atom, One of the Most Important Tools of Modern Physics: SO(4) to SO(4,2), Background, Theory, and Use in Calculating Radiative Shifts. Symmetry 12, 1323(2020). In this paper, there are two typographical errors: the right side of Eq. 297 should have a plus sign not a minus sign; on the right side of Eq. 299 the integral should have a minus sign and the ln term should  be $+\delta_{LO}ln\frac{2}{(Z\alpha)^2}$.

\bibitem{brow}Brown, L.S., Bounds on Screening Corrections in Beta Decay, Phys. Rev. 135, B314(1964).

\bibitem{bandsbook}Bethe, H.; Salpeter, E., \emph{The Quantum Mechanics of One and Two Electron Atoms}; Springer-Verlag,Berlin, Germany, 1957. 

\bibitem{sakurai}Sakurai, J. J., Modern Quantum Mechanics, Addison-Wesley, New York, New York (1994). 
\bibitem{morse}Morse, P.; Feshbach, H.,Methods of Theoretical Physics, Vol. 2, McGraw-Hill Book Co.: New York, NY, USA, 1953.

    

 
 

\end{thebibliography}
\end{document}